\documentclass[12pt,letterpaper]{article}

\usepackage[latin1]{inputenc}
\usepackage{amsmath,amssymb}
\usepackage{amscd}
\usepackage{amsfonts}

\usepackage{hyperref}
\hypersetup{colorlinks=true,linkcolor=red,citecolor=blue}

\newcommand{\LIE}[2]{\mathcal{L}_{#1}{#2}}
\renewcommand{\a}{\alpha}
\renewcommand{\b}{\beta}

\newcommand{\m}{\mu}
\newcommand{\n}{\nu}

\renewcommand{\th}{\theta}

\newcommand{\Dl}{\Delta}

\newcommand{\F}{\mathcal{F}}
\newcommand{\U}{\mathcal{U}}

\newcommand{\diff}{\mathfrak{D}}

\newcommand{\pa}{\partial}

\newcommand{\tihalf}{\tfrac{i}{2}}   

\newcommand{\ox}{\otimes}           

\renewcommand{\.}{\cdot}            

\title{A note on the implemetation of Poincar\'e symmetry in noncommutative field theory}
\author{Ignacio Cortese\footnote{nachoc@nucleares.unam.mx}, and  J. Antonio Garc\'ia\footnote{garcia@nucleares.unam.mx}\\
Departamento de F\'{\i}sica de Altas Energ\'{\i}as,
\\Instituto de Ciencias Nucleares, \\ Universidad Aut\'onoma de
M\'exico,\\ Apdo. Postal 70-543, M\'exico D.F.04510, M\'exico}

\begin{document}
 
\maketitle

\begin{abstract}
We argue that Poincar\'e symmetry can be implemented in NCFT if we allow the parameter of noncommutitive deformation $\theta^{\mu\nu}$ to change as a two-tensor under the corresponding space-time symmetry. The implementation is consistent with the definition of $\theta^{\mu\nu}$ in terms of space-time coordinates and with the Moyal star product. Inspired from the standard definition of a variational symmetry  we found a universal way to correct the implementation of the Poincar\'e symmetry by a term proportional to the variation of  $\theta^{\mu\nu}$ in such a way that the new transformation define a symmetry of the theory. Finally we present as an example the case of NCYM theory and comment about the obstructions to implement generalized space-time symmetries in NCFT like conformal or diffeomorphism transformations.

\end{abstract}

\section{Introduction}
The implementation of space-time symmetries in noncommutative field theory has been discussed by many authors in the last years. Among the obstruction for such implementation we have a fundamental restriction: if we define noncommutativity through
 the deformed commutation relation 
\begin{equation}\label{eq:ncr}
[{x}^\mu,{x}^\nu]_\star=x^\mu\star x^\nu- x^\nu\star x^\mu=i\vartheta\theta^{\mu\nu},
\end{equation}
where $\theta^{\mu\nu}$ is a constant antisymmetric matrix and we perform an infinitesimal space-time transformation given by $x^\mu\to {x'}^\mu=x^\mu+\xi^\mu(x)$ with $\xi^\mu$ an arbitrary function, we have
\begin{equation}
[{x'}^\mu,{x'}^\nu]_\star=i\vartheta\theta^{\mu\nu} 
+ i\vartheta(\theta^{\rho\nu}\partial_\rho \xi^\mu +\theta^{\mu\rho}\partial_\rho \xi^\nu)+{O}(\xi^2).
\end{equation}
The invariance of the relation  (\ref{eq:ncr}) implies then $\theta^{\rho\nu}\partial_\rho \xi^\mu +\theta^{\mu\rho}\partial_\rho \xi^\nu=0$. This restriction implies in turn a violation of Lorentz symmetry \cite{AlvarezGaume:2003}. We have then  different perspectives to work with symmetries in noncommutative field theories: To accept from the beginning  that Lorentz symmetry is lost with all the consequences that this breakdown carry on the formulation of a given field theory or to explore other alternatives like  the possibility that $\theta^{\mu\nu}(x)$ depend on the space-time point. Other  approaches are based on the change of the very definition of space-time symmetry transformation in NCFT.
 
The noncommutative parameter $\theta^{\mu\nu}$ can be considered as a matrix with numeric fixed entries or as a two tensor under the space-time transformation. In each case the physical interpretation of the ``coordinates'' $x^\mu$ changes.
An example of consistent implementation of (\ref{eq:ncr}) in  field theory is the so called Snyder space \cite{Sn1,Sn2} where the Lorentz symmetry is preserved. The price to pay for the explicit Lorentz invariance is that $\theta^{\mu\nu}$ is not anymore a constant matrix but instead it is related with the generators of the Lorentz group $J^{\mu\nu}$.

The first attempt to show by hand that in fact it is possible to implement the invariance of  NCFT under Poincar\'e transformations using functional derivatives and standard tools of variational calculus was given in \cite{BiGrGrKrPoScWu}.
But the implementation of space-time symmetries in NCFT is a delicate program. The action of symmetries in a noncommutative space-time is plagued with problems of consistence as the question about if the product that define the new deformed algebra ($\star$ product) change or not under the proposed space-time symmetry. In this work we will restrict ourselves to transformations that do not change the $\star$ product. Other open problems are the meaning of the integral measure in Moyal space, the implementation of variational calculus in noncommutative space-time, the consistent deformation of gauge transformations and the gauge fixing procedure.  In the last years a renewed interest in the search of different ways to define transformation rules that implement symmetries of NCFT have produced some important advances. Perhaps the better idea that we have at hand until now is the Drinfeld twist applied to the above deformation of standard Field Theories. As a result we have now a new concept of space-time symmetries called {\em twist} symmetries \cite{ChKuNiTu, Oeckl}. One of the nice properties of these symmetry transformations is that we can use them to implement Lorentz transformations in a given NCFT.

 Twisted  space-time symmetries was used to implement diffeomorphisms in noncommutative versions of Einstein gravity. We are still too far for a clear understanding of how these symmetries apply to gauge theories and/or curved space Field Theories. In these cases we have  good reasons to say that they are not consistent symmetries, at least if we insist in the Moyal deformation with a constant $\theta$.  Among the problems that arise when we try to implement twist symmetries in curved space we can mention that in order to define a co-product in the associated Hopf algebra of the symmetry algebra we need translation operators $P_\mu$ that commute among themselves but if we have a group manifold that define a gauge transformation or a curved space-time these translation operators must be changed by covariant derivatives that in general do not commute among themselves. In that instances (gauge or curved space-time) all that we can do is to construct non associative algebras. A different proposal for the construction of a consistent noncommutative Einstein gravity is \cite{CoGa} where we have implemented the symmetry under translations following the rules of this note.

At least for the simple case of global space-time symmetries the situation is under relative control. Nevertheless we still have some open problems in  our understanding of this relatively more easy problem. One is the implementation of the Noether theorem for this class of symmetries and the analysis of the associated Noether currents. Another important problem is that twist symmetries are very difficult to implement as variational symmetries because we do not have a corresponding variational approach in algebras where the co-product is not the trivial one. Recent attempts to overcome and/or try to understand better these issues are 
\cite{AsCaDi, Duenas, Chaichian:2010}

The formulation of local (gauge) symmetries on flat noncommutative space-time 
is also a delicate issue. Most gauge groups can not be defined on noncommutative space-time, 
because they do not close under the $\star$-product. Nevertheless the noncommutative unitary group $U_\star (N)$ 
can be defined in a consistent way. Its representations, however, are limited by a  no-go theorem \cite{ChPrSh-JaTu1, Chaichian:2009,Mat}.

In the general case (for a different gauge group) the associated enveloping algebra could have infinite gauge parameters and infinite gauge fields. Nevertheless and by fortune there exist a field redefinition from the original NCFT to a new effective theory which allow us to show that in fact there are not too many parameters and fields 
because we can relate the coefficients of the generators of the enveloping algebra in such a way that we have the correct number of gauge parameters corresponding to the given gauge group\footnote{A consistent truncation of the gauge enveloping algebra is also possible \cite{Barnich:2002}.}. In this way we can extend the possible noncommutative gauge groups to include special unitary groups as well: 
indeed, noncommutative gauge theories with gauge fields valued in the enveloping algebra 
of $su(n)$ has been constructed and a corresponding noncommutative version of the 
Standard Model has been worked out \cite{CaJuScWeWo,DoFrRo1,DoFrRo2}.  Also, using this approach to NCFT we can study all the space-time symmetries of Noncommutative Yang-Mills theory in flat Minkowski space. In four dimensions the special conformal transformations and dilatations are obstructed. The Poincar\'e transformations are also restricted to the subgroup that preserves the constant matrix $\theta^{\mu\nu}$ \cite{BaBrGr}. 

The Seiberg-Witten map can be constructed from
a modern point of view, where NCFT is considered as a consistent deformation of standard commutative field theory or from low energy limits of string theories with modified backgrounds \cite{SeWi}. This map provide another approach to the noncommutative gauge theories. In its simplest presentation it relate a noncommutative $U_\star (N)$ gauge theory to 
a commutative one, obtained as low-energy effective limit in string theory, by using two 
different regularization methods (the point-splitting method and the Pauli-Villars method, 
respectively).

The fact that a given NCFT can be mapped to a commutative effective field theory with infinite new vertices open  new questions related with the implementation and significance of symmetries and conserved currents in each side of this mapping. 
In particular problems like what is the relation, if any, between a twisted version of a space time symmetry on the noncommutative theory with the standard space time symmetry ? It is possible to define a conserved current (symmetry) in the commutative side and then return to the noncommutative theory using the SW field redefinition and obtain the corresponding noncommutative version of the conserved current (symmetry)?    
Is the resulting symmetry a twisted symmetry? As we know, symmetries and its associated conserved currents differ at most by trivial symmetries (conserved currents) after field redefinitions\footnote{For a modern presentation of the Noether theorem in FT see \cite{Barnich:1994}.}. We can naively expect  that the standard symmetries (conserved currents) will work on either side, again up to trivial symmetries (conserved currents). Unfortunately this is not the case and this is an interesting open problem. The associated problems with gauge transformations like the implementation of the Dirac algorithm for constrained theories, the relation between first class constraints and gauge symmetries and the gauge fixing procedure are also interesting open problems in NCFT.  

As a first step toward an answer to these questions we propose a consistent implementation of Poincar\'e symmetry in NCFT based on previous experiences on twist and variational symmetries. We will work with the effective description of the NCFT given after the SW map. This theories are formulated in terms of standard fields (as opposed to noncommutative fields) and where the usual technics of variational derivatives can be used.
Our approach is classical and its goal is to have a better understanding of symmetries, variational principles and eventually a Noether theorem in NCFT. 

We want to stress here the fact that it is possible to construct a consistent NCFT with manifest Lorentz invariance allowing that the $\partial_\mu$ transform as space-time vectors and $\theta^{\mu\nu}$ transform as a two-tensor. In this way the $\star$ product is invariant. We can then compare this analysis with the variational and twist implementation of symmetries. In our setup $\theta$ itself is not invariant but quantities like $\theta^{\mu\nu}\theta_{\mu\nu}, p_\mu\theta^{\mu\nu}k_\nu$ are Lorentz invariant so we can still have a fundamental ``scale'' and an $S$ matrix that could be Lorentz invariant.
In the next section we will review the definition of twist symmetries from a physical motivated setup. In section 3 we will present our central idea: a proposal of a modified variational implementation of noncommutative symmetries in NCFT. Here we will study the transformation properties of the $\star$ product under space-time mappings showing that they are severely restricted if we want invariance of the $\star$ product under space-time transformations. We will also present the variational implementation for the allowed space-time symmetries and relate our result with the one obtained using twist symmetries. In section 4 we work the example of NCYM theory and show how the implementation of Poincar\'e symmetry can be done. The last section 5 is devoted to discussion and comments about the relation of our work with other works whose aim was to implement also consistent space-time symmetries in NCFT.

As a final comment we like to mention other structural  problems in the formulation of NCFT like the UV/IR mixing, causality,  unitary and non-locality. In particular, the UV/IR mixing ruins the possibility to apply renormalization theory in NCFT. For a recent discussion about this interesting topic see \cite{Blaschke:2010}.

\section{Twisted space-time symmetries defined from covariance of the star-product}

In the following we will remind the reader with the basics of a physical motivated introduction to the concept of twisted symmetry. For details see \cite{Al-GaMeVa-Mo}. We present also the notation and some relevant aspects of the standard derivation of twisted symmetries from Hopf algebras. 

Consider a space-time diffeomorphism generated by a vector field $\xi(x)=\xi^\mu\partial_\mu$ and two fields $\Phi_1,\Phi_2$ transforming according to its own finite representation ${\cal R}_1$, ${\cal R}_2$ given explicitly in terms of some differential operators
$$\Phi_{1,2}'={\cal D}_{1,2}(\xi)\Phi_{1,2}$$
Denoting as $m$ the ordinary product of fields acting on the associated bi-algebra and by $m_\star$ the corresponding $\star$ product we can write the definition
 \begin{equation*}
(\phi\star\psi)(x)=m_\star(\phi\otimes\psi)(x)= m[\mathcal{F}^{-1}(\phi\otimes\psi)](x) 
\end{equation*}
where
\begin{equation}
\label{fcal}
\mathcal{F}^{-1}\equiv \exp{(\frac{i\vartheta}{2}\theta^{\alpha\beta}\partial_\alpha\otimes\partial_\beta)} \\ =\sum\limits^{\infty}_{n=0}\frac{(i\vartheta/2)^n}{n!}\theta^{\mu_1\nu_1}\cdots\theta^{\mu_n\nu_n}\left(\partial_{\mu_1}\cdots\partial_{\mu_n}\otimes\partial_{\nu_1}\cdots\partial_{\nu_n}\right).
\end{equation}
We will require a basic condition on the transformation properties of the $\star$ product of two fields that in fact can be considered as the definition of the action of a twisted diffeomorphism: 
\begin{center}
\begin{equation}\label{basic}
\vbox{\em The $\star$ product of the two fields $\Phi_1\star\Phi_2$ transform
in the product of the above representations ${\cal R}_1\otimes {\cal R}_2$.}  
\end{equation}
\end{center}
i.e, we require that the $\star$ product of two tensor fields transform 
covariantly,  
${T_1}^{(p_1)}_{(q_1)}(x)\star {T_2}^{(p_2)}_{(q_2)}(x)$  transform as 
a tensor of type $(p_1 
+ p_2 , q_1 + q_2)$\footnote{A twist for a vector space $V$ is defined as a map $\mathcal{T}:V\otimes V\to V\otimes V$ such that for $v_i$ a base in $V$, $v_i\otimes v_j\mapsto\mathcal{T}(v_i\otimes v_j)= M^{kl}_{ij}v_k\otimes v_l$.
}. This can be achieved 
by imposing (\ref{basic}) in the definition of the star-product of the two given fields,
\begin{equation}
({\cal F}^{-1}\Phi_1\otimes\Phi_2)'=[{\cal D}_1(\xi)\otimes{\cal D}_2(\xi)]({\cal F}^{-1}\Phi_1\otimes\Phi_2)={{\cal F}'}^{-1}\Phi'_1\otimes\Phi'_2
\end{equation}
which implies in turn that $\cal F$ changes under the diffeomorphism as
\begin{equation}\label{changeF}
{{\cal F}'}^{-1}=[{\cal D}_1(\xi)\otimes{\cal D}_2(\xi)]{\cal F}^{-1}[{\cal D}_1^{-1}(\xi)\otimes{\cal D}^{-1}_2(\xi)]
\end{equation}
The infinitesimal version of (\ref{changeF}) is
$$\delta_\xi{\cal F}^{-1}={\cal F}^{-1}\sum_{n=1}^\infty\frac{(-i\vartheta/2)^n}{n!}\theta^{\mu_1\nu_1}...\theta^{\mu_n\nu_n}\{[\partial_{\mu_1},[\partial_{\mu_2},...[\partial_{\mu_n},\delta_{\xi,1}]...]]\otimes\partial_{\nu_1}\partial_{\nu_2}...\partial_{\nu_n}$$
$$+\partial_{\mu_1}\partial_{\mu_2}...\partial_{\mu_n}\otimes[\partial_{\nu_1},[\partial_{\nu_2},...[\partial_{\nu_n},\delta_{\xi,2}]...]]\},$$
where $\delta_{\xi,1,2}$ are the infinitesimal versions of the differential operators ${\cal D}_{1,2}$. Therefore the transformation of the star-product of $\Phi_1$ and $\Phi_2$ is given by

$$\delta_\xi(\Phi_1(x)\star\Phi_2(x))=m[{{\cal F}'}^{-1}\Phi'_1\otimes\Phi'_2]-m[{\cal F}^{-1}\Phi_1\otimes\Phi_2]$$
\begin{equation}\label{act*}
=\delta_{\xi,1}\Phi_1(x)\star\Phi_2(x)+\Phi_1(x)\star \delta_{\xi,2}\Phi_2(x)+\Phi_1(x)(\delta_{\xi}\star)\Phi_2(x)
\end{equation}
where
\begin{equation}
\Phi_1(x)(\delta_{\xi}\star)\Phi_2(x)=\sum_{n=1}^\infty\frac{(-i\vartheta/2)^n}{n!}\theta^{\mu_1\nu_1}...\theta^{\mu_n\nu_n}\{[\partial_{\mu_1},[\partial_{\mu_2},...[\partial_{\mu_n},\delta_{\xi,1}]...]]\Phi_1\star\partial_{\nu_1}\partial_{\nu_2}...\partial_{\nu_n}\Phi_2
\nonumber
\end{equation}
\begin{equation}
\label{noleib}
+\partial_{\mu_1}\partial_{\mu_2}...\partial_{\mu_n}\Phi_1\star[\partial_{\nu_1},[\partial_{\nu_2},...[\partial_{\nu_n},\delta_{\xi,2}]...]]\Phi_2\}
\end{equation}
As a result of the deformation the infinitesimal transformations are not acting as derivations (Leibniz rule) on the $\star$ product of two fields. In fact we can read this last relation as a change in the $\star$ product given by the action of the space-time transformation. But notice the important fact that the change in the $\star$ product is an induced change by the transformation properties of the fields. As we can not change the algebra itself under the action of a symmetry group because the concept of symmetry as an homomorphism of the algebra of functions is meaningless, we prefer to adopt this second perspective and read the relation (\ref{noleib}) as an induced change that comes from the definition of the transformation properties of the fields. In this way we can see that this result exactly matches the one obtained by the deformation of the coproduct of the associated Hopf algebra. 
 Indeed, the deformation of the coproduct  can be read from the transformation induced on the star-product by the requirement (\ref{basic}). But notice that the construction presented here is self contained and is independent of the algebraic construction based on the twist of a given Hopf algebra. 

Let us remark that the result (\ref{act*}) codifies the action of a  space-time transformation on the product of two fields. The action of the transformation on the fields is not changed and corresponds  to the usual one. In particular Poincar\'e transformations act in the usual way on fields but are deformed when they act on the $\star$ product of two or more fields. It is a nice implementation of the transformation if we have present that
the $\star$ product itself does not change under the given diffeomorphism. This last statement  is just a postulate that we need for the general consistence of the deformation procedure. The usual way to implement the invariance of the star product is to enforce the condition that $\theta^{\mu\nu}$ and the space-time partial derivatives do not change under the diffeomorphism $\xi$. In particular if $\Phi$ transform in the representation ${\cal R}$ then $\partial_{\mu_1}....\partial_{\mu_n}\Phi$ transform under the same representation as $\Phi$. This may sound to radical but in fact is what we do when  work with twisted symmetries: the noncommutative parameter $\theta$ is a constant matrix (not a (0,2)-tensor !) and the space-time coordinates  $x^\mu$ are scalars under the diffeomorphism in order to be consistent with the definition of $\theta$ given in (\ref{eq:ncr}). Under these assumptions it is easy to find that $\delta_\xi\theta^{\mu\nu}=\frac{1}{i\vartheta}\delta_\xi([x^\mu,x^\nu]_\star)=0$.

In particular, using this rules we can construct NCFT invariant under the twisted Poincar\'e algebra. As a bonus the symmetry algebra is not deformed when acting on primary fields (the field content of the NCFT) and it can be used for the standard definition and classification of particles according to the finite representations of  the underlying symmetry algebra. What we are doing is changing the coproduct of the algebra but  keeping the same rules for the Lie product between its generators. It is not difficult to check  the twisted invariance of a standard NCFT, constructed in the usual way from a given FT, if the original QFT already have the corresponding non twisted invariance. What remains quite non-trivial is the problem of the construction of the associated conserved currents and the construction of the corresponding variational implementation of this type of symmetries. It is not clear if the twisted symmetries constructed in this way can be used as symmetry principles to construct consistent NCFT's but nevertheless they can be considered as a first step to achieve that goal.

The analogous presentation of the ideas of this section in the context of Hopf algebras can be resumed as follows: 
Consider the Lie algebra $\diff$ of diffeomorphisms, whose
generators are vector fields.
Take as the Hopf algebra $H$ the enveloping algebra~$\U(\diff)$.
Likewise the enveloping algebra of any Lie algebra, the coproduct
$\Dl$ is first defined for elements $h$ of $\diff$ by $\Dl(h)=1\ox
h+h\ox1$, and then multiplicatively extended to all of $\U(\diff)$ by
means of $\Dl(hh')=\Dl(h)\Dl(h')$. Now consider the algebra of functions on
spacetime with the ordinary multiplication $m(f\ox g)=fg$ and $\F=\exp(-\tihalf\,\th^{\m\n}\pa_\m\ox\pa_\n )$. This $\F$
is clearly in $\U(\diff)\ox\U(\diff)$, and has an inverse
\begin{equation*}
 \F^{-1} = \exp(\tihalf\,\th^{\m\n}\pa_\m\ox\pa_\n)
\end{equation*}
The Moyal product is defined as
\begin{equation}
   \mu (f\ox g) = m\big( \F^{-1} \.(f\ox g)\,\big)
      = f\star g\,.
\label{eq:twisted-product}
\end{equation}
 For the generators of translations, Lorentz
transformations and dilatations~\cite{Matlock} the
following expressions were obtained, 
\begin{align}
   \Dl_T(P_\m) & = P_\m\ox 1 + 1\ox P_\m  \nonumber\\[3pt]
   \Dl_T(M_{\m\n}) & = M_{\m\n} \ox 1 + 1 \ox M_{\m\n}
       \nonumber\\
   & + \tihalf\,\th^{\a\b}
           \big[ (g_{\m\a}P_\n - g_{\n\a}P_\m) \ox P_\b
                + P_\a \ox (g_{\m\b}P_\n -  g_{\n\b}P_\m)\big]
       \label{eq:complicated} \\[3pt]
   \Dl_T(D) & = D \ox 1 + 1 \ox D - i\,\th^{\m\n} P_\m \ox P_\n\,.
      \nonumber
\end{align}
These equations precisely show that $\Dl_T(M_{\m\n})$ and
$\Dl_T(D)$ are not derivations of the Moyal product. From
eq.~\eqref{eq:complicated} it was concluded that the Poincar\'e group
remains relevant in noncommutative field theory. 
 The generators $K_\mu$ of special conformal
transformation could be added to the list of
computed~$\Dl_T(h)$~\cite{Matlock}. Now, because we are in the
enveloping algebra, the method applies to
differential operators of any order. The method is thus a recipe to
encode the action of arbitrary differential operators with polynomial
coefficients on Moyal products.

The previous remark leads in a systematic and simple way to compute
the twisted coproduct of the generator of any spacetime
transformation. Let us take an infinitesimal spacetime transformation
generated by differential operators of the form $x^{\m_1}\cdots
x^{\m_N}\pa_\n$. In particular we can calculate
\begin{equation*}
   \mu\big( \Dl_T(x^{\m_1}\cdots x^{\m_N}\pa_\n)\.
     (x^\a\ox x^\b - x^\b \ox x^\a)\big) = 0\,.
\end{equation*}
where the space time coordinates are explicitly considered as scalars under the given diffeomorphism. This calculation imply that $\th^{\a\b}$ remains unchanged as we expected from the definition of the twist symmetries.

\section{Modified Variational Implementation of space-time symmetries}

From the definition of symmetry as a map of the solution space onto itself of a given field theory we require that the algebra of functions does not change. So as a first step to implement symmetries in a noncommutative space-time  we will explore in this section the transformation properties of the star product under space-time symmetries of the flat space-time metric, i.e., under the transformation generated by the killing vectors of $\eta_{\mu\nu}$.  
A central point in our analysis is that $\theta^{\mu\nu}$ will be considered as a $(2,0)$-tensor under the space-time transformations in contrast with the twisted version of the corresponding transformation. We will see, using standard tools of variational calculus in field theory that its is possible to give a consistent implementation of space-time transformations in NCFT if we can find the subset of the symmetries of a given background metric that leave invariant the $\star$ product.

In $d$ dimensions the symmetries of the flat space-time metric satisfies the killing equation 
\begin{equation}\label{killingeq}
\partial_\mu\xi_\nu+\partial_\nu\xi_\mu=\frac 2d\eta_{\mu\nu}\partial_\sigma\xi^\sigma
\end{equation}
The general solution to this equation is
$$\xi^\mu=a^\mu+\lambda^\mu_\nu x^\nu+cx^\mu+b^\nu\eta_{\nu\rho}x^\rho x^\mu-\frac12 b^\mu\eta_{\nu\rho}x^\nu x^\rho$$
corresponding to translations $a^\mu$, Lorentz transformations $\lambda_{\mu\nu}$, dilatations $c$ and special conformal transformations $b^\mu$. 

If we restrict ourselves to the context where $\theta^{\mu\nu}$ is a constant only the affine subgroup can be considered. 
The reason is that an infinitesimal change in $\theta^{\mu\nu}$ given by any of the killing vectors of the flat metric have the generic form
\begin{equation}
\label{lietheta}
\delta\theta^{\mu\nu}=-\LIE{\xi}{\theta^{\mu\nu}}=-\xi^\rho\partial_\rho\theta^{\mu\nu}+\partial_{\rho}\xi^{\mu}\theta^{\rho\nu}+\partial_{\rho}\xi^{\nu}\theta^{\mu\rho}.
\end{equation}
If $\theta'^{\mu\nu}=\theta^{\mu\nu}+\delta\theta^{\mu\nu}$, 
it is easy to see that only the affine transformations $x^\mu \mapsto x'^\mu = x^\mu + \xi^\mu$, with $\xi^\mu={B^\mu}_\nu x^\nu + a^\mu$ (where ${B^\mu}_\nu$ and $a^\mu$ are constants), leave $\theta'$ independent of the space-time point. Hence, only the Weyl transformations can be considered. Now we will show that the $\star$ product is also invariant under this subclass of symmetries of the background. 

We begin  by exploring the change in 
$\mathcal{F}^{-1}$ under an infinitesimal Weyl transformation. As $\mathcal{F}^{-1}$ is a bi-algebra operator and we need to extract a factor of the product of the two algebras in order to have a bi-algebra operator of the same form as the original one,  we conclude that the only allowed diffeomorphims from the general solution of the killing equation (\ref{killingeq}) are the terms at most linear in $x^\mu$.  For this subclass of killing vectors (affine or Weyl transformations) we will show that $\mathcal{F}^{-1}$ remains in fact invariant. The change in  $\mathcal{F}^{-1}$ is
\begin{multline*}
\mathcal{F}'^{-1}\equiv \exp{(\frac{i\vartheta}{2}\theta'^{\alpha\beta}\partial'_\alpha\otimes\partial'_\beta)} \\ =\sum\limits^{\infty}_{n=0}\frac{(i\vartheta/2)^n}{n!}\theta'^{\mu_1\nu_1}\cdots\theta'^{\mu_n\nu_n}\left(\partial'_{\mu_1}\cdots\partial'_{\mu_n}\otimes\partial'_{\nu_1}\cdots\partial'_{\nu_n}\right),
\end{multline*}
A generic term of order $n$ in $\vartheta$ is
\begin{multline*}
\theta'^{\mu_1\nu_1}\cdots\theta'^{\mu_n\nu_n}=(\theta^{\mu_1\nu_1}+\delta\theta^{\mu_1\nu_1})\cdots(\theta^{\mu_n\nu_n}+\delta\theta^{\mu_n\nu_n}) \\ 
=\theta^{\mu_1\nu_1}\cdots\theta^{\mu_n\nu_n}+\delta(\theta^{\mu_1\nu_1}\cdots\theta^{\mu_n\nu_n}) \\
=\theta^{\mu_1\nu_1}\cdots\theta^{\mu_n\nu_n}+\sum\limits_{i=1}^n\theta^{\mu_1\nu_1}\cdots\delta\theta^{\mu_i\nu_i}\cdots\theta^{\mu_n\nu_n},
\end{multline*}
so the product of $n$ partial derivatives can be written as\footnote{With ${\partial'}_\mu=\frac{\partial x^\nu}{{\partial x'}^\mu}\partial_\nu$ y $\frac{\partial x^\nu}{{\partial x'}^\mu}=\delta^\nu_\mu-\partial_\mu\xi^\nu$.}
\begin{multline}
\label{transpar}
\partial'_{\rho_1}\cdots\partial'_{\rho_n}=(\delta_{\rho_1}^{\sigma_1}-\partial_{\rho_1}\xi^{\sigma_1})\partial_{\sigma_1}\cdots(\delta_{\rho_n}^{\sigma_n}-\partial_{\rho_n}\xi^{\sigma_n})\partial_{\sigma_n} \\
=(\delta_{\rho_1}^{\sigma_1}-\partial_{\rho_1}\xi^{\sigma_1})\cdots(\delta_{\rho_n}^{\sigma_n}-\partial_{\rho_n}\xi^{\sigma_n})\partial_{\sigma_1}\cdots\partial_{\sigma_n} \\
=\left(\delta_{\rho_1}^{\sigma_1}\cdots\delta_{\rho_n}^{\sigma_n}-\sum\limits_{i=1}^n\delta_{\rho_1}^{\sigma_1}\cdots(\partial_{\rho_i}\xi^{\sigma_i})\cdots\delta_{\rho_n}^{\sigma_n}\right)\partial_{\sigma_1}\cdots\partial_{\sigma_n},
\end{multline}
where $\partial_\rho\xi^\sigma$ is a constant. The final result is
\begin{align}
\label{thetapar}
&\theta'^{\mu_1\nu_1}\cdots\theta'^{\mu_n\nu_n}\partial'_{\mu_1}\cdots\partial'_{\mu_n}\otimes\partial'_{\nu_1}\cdots\partial'_{\nu_n} = \theta^{\mu_1\nu_1}\cdots\theta^{\mu_n\nu_n}\partial_{\mu_1}\cdots\partial_{\mu_n}\otimes\partial_{\nu_1}\cdots\partial_{\nu_n} \notag \\
&-\sum\limits_{i=1}^n
[\theta^{\mu_1\nu_1}\cdots\theta^{\mu_n\nu_n}(\delta_{\mu_1}^{\alpha_1}\cdots\delta_{\mu_n}^{\alpha_n}\delta_{\nu_1}^{\beta_1}\cdots\partial_{\nu_i}\xi^{\beta_i}\cdots\delta_{\nu_n}^{\beta_n} \notag \\
&\hskip2in +\delta_{\mu_1}^{\alpha_1}\cdots\partial_{\mu_i}\xi^{\alpha_i}\cdots\delta_{\mu_n}^{\alpha_n}\delta_{\nu_1}^{\beta_1}\cdots\delta_{\nu_n}^{\beta_n}) \notag \\
&\qquad -\theta^{\mu_1\nu_1}\cdots\delta\theta^{\mu_i\nu_i}\cdots\theta^{\mu_n\nu_n}\delta_{\mu_1}^{\alpha_1}\cdots\delta_{\mu_n}^{\alpha_n}\delta_{\nu_1}^{\beta_1}\cdots\delta_{\nu_n}^{\beta_n}]\partial_{\mu_1}\cdots\partial_{\mu_n}\otimes\partial_{\nu_1}\cdots\partial_{\nu_n}.
\end{align}
The $i$-th term in this expression can be written as
\begin{multline*}
\theta^{\mu_1\nu_1}\cdots\theta^{\mu_{i-1}\nu_{i-1}}\Bigl[\delta_{\mu_1}^{\alpha_1}\cdots\delta_{\mu_{i-1}}^{\alpha_{i-1}}\delta_{\nu_1}^{\beta_1}\cdots\delta_{\nu_{i-1}}^{\beta_{i-1}}\Bigl(\theta^{\mu_i\nu_i}\delta_{\mu_i}^{\alpha_i}\partial_{\nu_i}\xi^{\beta_i}+\theta^{\mu_i\nu_i}\partial_{\mu_i}\xi^{\alpha_i}\delta_{\nu_i}^{\beta_i}  \\
 -\delta\theta^{\mu_i\nu_i}\delta_{\mu_i}^{\alpha_i}\delta_{\nu_i}^{\beta_i}\Bigr)\delta_{\nu_{i+1}}^{\beta_{i+1}}\cdots\delta_{\nu_n}^{\beta_n}\delta_{\mu_{i+1}}^{\alpha_{i+1}}\cdots\delta_{\mu_n}^{\alpha_n}\Bigr]\theta^{\mu_{i+1}\nu_{i+1}}\cdots\theta^{\mu_n\nu_n}.
\end{multline*}
Using
\begin{equation}
\label{deltatheta1}
\theta^{\alpha_i\nu_i}\partial_{\nu_i}\xi^{\beta_i}+\theta^{\mu_i\beta_i}\partial_{\mu_i}\xi^{\alpha_i} -\delta\theta^{\alpha_i\beta_i}=0, 
\end{equation}
and from (\ref{thetapar}) we have   $\mathcal{F}'^{-1}=\mathcal{F}^{-1}$ so we conclude that $\star'=\star$. 

The equation (\ref{deltatheta1}) coincide with the definition of the transformation of $\theta^{\mu\nu}$ as a constant (2,0)-tensor. We conclude that if we transform in a  covariant way  $\theta^{\mu\nu}$ and the partial derivatives, the $\star$ product is invariant under Weyl transformations.  Unfortunately our argument does not apply in the case of conformal transformations. Even though  $\theta$ is different in each reference frame, we can construct invariants using $\theta^{\mu\nu}$ and the background metric. For example $\eta_{\mu\alpha}\eta_{\nu\beta}\theta^{\mu\nu}\theta^{\alpha\beta}$ could be a good observable quantity in a given NCFT.

With the previous ideas in mind, we wonder if it is possible to construct a variational implementation of space-time transformations that captures the basic features of the twisted transformations but at the same time transform $\theta^{\mu\nu}$ as a two tensor.
Lets consider the twist transformation corresponding to $x^\mu\mapsto x^\mu +\xi^\mu$ with $\xi^\mu=B^\mu_\nu x^\nu$ and $\delta_\xi\phi_{1(2)}=-\xi^\rho\partial_\rho \phi_{1(2)}$ . From (\ref{noleib}) we have
\begin{equation*}
\phi_1(x)(\delta_{\xi}\star)\phi_2(x)=\frac{(-i\vartheta)}{2}\theta^{\mu\nu}\Bigl([\partial_{\mu},\delta_{\xi,1}]\phi_1\star\partial_{\nu}\phi_2+\partial_{\mu}\phi_1\star[\partial_{\nu},\delta_{\xi,2}]\phi_2\Bigr)+\mathcal{O}(\vartheta^2),
\end{equation*}
where $[\partial_\mu , \delta_{\xi,1(2)}]\phi_{1(2)}=-(\partial_\mu\xi^\rho)\partial_\rho\phi_{1(2)}=-{B^\rho}_\mu\partial_\rho\phi_{1(2)}$. 
It is easy to see that if  $\xi^\mu$ is linear in the coordinates,  the higher order terms in $\vartheta$ are zero. So the twisted rule for the implementation of a linear transformation is
\begin{multline}
\label{twsitasvartheta}
\delta_\xi(\phi_1(x)\star\phi_2(x))=(-\xi^\rho\partial_\rho\phi_1)\star\phi_2+\phi_1\star(-\xi^\rho\partial_\rho \phi_2)\\ 
+i\frac\vartheta2\Bigl(\theta^{\rho\nu}{B^\mu}_\rho +\theta^{\mu\rho}{B^\nu}_\rho\Bigr)\partial_\mu\phi_1\star\partial_\nu\phi_2.
\end{multline}
What we want to remark here is that the last term in this result can be obtained also from the variation $\phi_1\star\phi_2$ with respect to $\theta^{\mu\nu}$. Notice that $\delta\theta^{\mu\nu}\equiv -\LIE{\xi}{\theta^{\mu\nu}}$ so we can define
\begin{equation*}
\delta_\theta(\phi_1\star\phi_2)\equiv \frac12(\delta\theta^{\mu\nu})\frac{\partial(\phi_1\star\phi_2)}{\partial \theta^{\mu\nu}}=i\frac\vartheta2\Bigl(\theta^{\rho\nu}{B^\mu}_\rho +\theta^{\mu\rho}{B^\nu}_\rho\Bigr)\partial_\mu\phi_1\star\partial_\nu\phi_2,
\end{equation*}
where $\frac{\partial\theta^{\alpha\beta}}{\partial\theta^{\mu\nu}}=\delta^\alpha_\mu\delta^\beta_\nu-\delta^\alpha_\nu\delta^\beta_\mu$. As a consequence of this result we will see that it is possible to implement in a universal way the Weyl group in a given NCFT if at the same time we vary with respect to $\theta^{\mu\nu}$. If the Weyl group is a symmetry of the undeformed action then the resulting prescription that we will construct here will be a symmetry of the corresponding NCFT.

Lets consider the  standard variational implementation of space-time symmetries in field theory and the killing vector
 $\xi^\mu=a^\mu+{\lambda^\mu}_\nu x^\nu+cx^\mu= a^\mu+{B^\mu}_\nu x^\nu$, with ${B^\mu}_\nu\equiv {\lambda^\mu}_\nu + c\delta^\mu_\nu$. The variational derivative along $\xi^\mu$ is
\begin{equation}
\label{dervar}
W_\xi^{\phi}\equiv\delta_\xi\phi_a (y)\frac{\delta}{\delta\phi_a (y)},
\end{equation}
with $a$ an internal or space-time index. This variational implementation of the symmetry along $\xi^\mu$ acts on the $\star$ product as 
\begin{equation*}
W_\xi^{\phi} (U \star V)=(W_\xi^{\phi} U)\star V +U\star(W_\xi^{\phi} V).
\end{equation*}
In NCFT the problem with this implementation of the variational derivative is that the basic principle given by (\ref{basic}) is violated.  For example if we have two scalar fields $\phi_1$ y $\phi_2$  then
\begin{multline*}
W_\xi^{\phi}[\phi_1(x)\star\phi_2(x)]=(W_\xi^{\phi}\phi_1(x))\star\phi_2(x)+\phi_1(x)\star(W_\xi^{\phi} \phi_2(x)) \\ =-\xi^\alpha\partial_\alpha(\phi_1(x)\star\phi_2(x))-i\frac\vartheta2(\theta^{\rho\beta} B^\alpha_\rho+\theta^{\alpha\rho}B^\beta_\rho)\partial_\alpha\phi_1\star\partial_\beta\phi_2,
\end{multline*}
where we used the indentities 
\begin{align}
\label{regmult}
x^\alpha(\phi\star\psi)&=(x^\alpha\phi)\star\psi-\frac{i\vartheta}{2}\theta^{\alpha\beta}(\phi\star\partial_\beta\psi) \notag \\
&=\phi\star(x^\alpha\psi)+\frac{i\vartheta}{2}\theta^{\alpha\beta}(\partial_\beta\phi\star\psi).
\end{align}
This result imply that $\phi_1(x)\star\phi_2(x)$ does not transform as a scalar. 

Nevertheless we can modify the definition of the variational derivative by adding to it a term that takes into acount the variaton in  $\theta^{\mu\nu}$ in such a way that the new rule comply with the basic principle (\ref{basic}). The proposal is
\begin{equation}
\label{twistvariacional}
\boxed{W_\xi\equiv W_\xi^{\phi}+W_\xi^{\theta}},
\end{equation}
where the term $W_\xi^{\theta}$ acts on the product of two fields $U\star V$ as
\begin{equation}
\label{wtheta}
W_\xi^{\theta} (U \star V)=(W_\xi^{\theta} U)\star V +U\star(W_\xi^{\theta}V)+\frac{i\vartheta}{2}(\delta\theta^{\alpha\beta})(\partial_\alpha U\star\partial_\beta V),
\end{equation}
with $W_\xi^{\theta} U\equiv \frac12\delta\theta^{\alpha\beta}\frac{\partial U}{\partial \theta^{\alpha\beta}}$, and $\delta\theta^{\alpha\beta}\equiv-\LIE{\xi}{\theta^{\alpha\beta}}=\partial_{\rho}\xi^{\alpha}\theta^{\rho\beta}+\partial_{\rho}\xi^{\beta}\theta^{\alpha\rho}$.
It is easy to see that with this modification the prescription (\ref{twistvariacional}) respects the tensorial character of the fields. For example if $\hat{F}_{\mu\nu}=\partial_\mu \hat{A}_\nu-\partial_\nu \hat{A}_\mu -i[\hat{A}_\mu,\hat{A}_\nu]_\star$ then $W_\xi \hat{F}_{\mu\nu}=-\LIE{\xi}{\hat{F}_{\mu\nu}}$. We can now recover the action of symmetries in NCFT. The price to pay is that the prescription is not a derivation of the $\star$ product. The relation  (\ref{twistvariacional}) is a basic result of this note.

Comparing this result with twist symmetries defined in the previous section we can see that the implementations are analogous in the sense that they comply with the basic principle (\ref{basic}) but they are quite different because the prescription  (\ref{twistvariacional}) is variational and works with the fields of a given theory in such a way that every space-time index is transformed in the usual way. In contrast the twist symmetries acting on the $\star$ product of fields of two or more fields
do not transform $\theta^{\mu\nu}$ which is treated as a matrix. 
It is important to remark that the implementation (\ref{twistvariacional}) only works for Weyl symmetries. In this case the twist symmetries or the prescription (\ref{twistvariacional}) give the same result and we can conclude that the correspondence of the twist symmetry after the Seiberg-Witten map is the modified variational symmetry given here.
A recent approach along the same lines are  \cite{Pi1,Pi2}.

\section{An example: Weyl invariance of Noncommutative Yang-Mills Theory (NCYM)}

For a given NCFT with gauge invariance we have two different approaches to study the effects of a symmetry in the fields and in the corresponding noncommutative action: the noncommutative theory in the Moyal space or the equivalent theory in terms of commutative algebra after de Seiberg-Witten map. In this section we will focus  on this second approach.

Using this map the NCYM theory can be written in terms of an effective Lagrangian with infinite  local vertices and the standard local symmetry of the Yang-Mills theory. If we denote by $\hat{A}_\mu^B\equiv f_\mu^B([A],\theta;\vartheta)$ as the effect of the Seiberg-Witten map on the gauge fields, then NCYM theory can be written as
\begin{equation}
\label{effNCYM}
S_{YMNC}=-\frac{1}{4}\int (dx)^4 \mathfrak{tr}\Bigl[\hat{F}_{\mu\nu}\hat{F}^{\mu\nu}\left(\hat{A} = f([A],\theta;\vartheta)\right)\Bigr]\equiv\int (dx)^D{L}^{eff}([A],\theta;\vartheta) ,
\end{equation}
To leading order in the deformation parameter  $\vartheta$ the effective Lagrangian is \cite{BaBrGr}
\begin{multline}
\label{effNCYM-1}
S_{YMNC}=\int (dx)^4({L}^{ef(0)}([A],\theta)+{L}^{ef(1)}([A],\theta;\vartheta)+O(\vartheta^2) )\\\equiv-\frac{1}{4}\int (dx)^D \mathfrak{tr}\Bigl(F_{\mu\nu}F^{\mu\nu}+\frac{i\vartheta\theta^{\alpha\beta}}{2}(-F_{\alpha\beta}F_{\mu\nu}F^{\mu\nu}+4F_{\alpha\mu}F_{\beta\nu}F^{\mu\nu})+O(\vartheta^2)\Bigr).
\end{multline}
The next to leading order terms are in general complicated expressions that include $A_\rho, F_{\mu\nu}$ and its derivatives \cite{Mo,UlYa}. Nevertheless, we we can write the effect of the Seiberg-Witten map of $\hat{F}_{\mu\nu}$ in the generic form
\begin{multline}
\label{Fexp}
\hat{F}_{\mu\nu}=F_{\mu\nu}+\vartheta\theta^{\alpha\beta}f_{\alpha\beta\mu\nu}+\vartheta^2\theta^{\alpha_1\beta_1}\theta^{\alpha_2\beta_2}f_{\alpha_1\beta_1\alpha_2\beta_2\mu\nu}+... \\ =\sum\limits_{n=0}^{\infty}\vartheta^{n}\theta^{\alpha_1\beta_1}...\theta^{\alpha_n\beta_n}f_{\alpha_1\beta_1\cdots\alpha_n\beta_n\mu\nu},
\end{multline}
where $f_{\alpha_1\beta_1 ...\alpha_n\beta_n\mu\nu}$ contain $A_\rho, F_{\mu\nu}$ and its derivatives\footnote{With the ``initial condition''  $\theta^{\alpha_1\beta_1}...\theta^{\alpha_n\beta_n}f_{\alpha_1\beta_1 ...\alpha_n\beta_n\mu\nu}\mid_{n=0}\equiv F_{\mu\nu}=\partial_\mu A_\nu - \partial_\nu A_\mu -i[A_\mu,A_\nu]$.} (but not $\theta$). Lets define for each $n=0,1,2,\dots$ the functions
\begin{equation}
\label{f}
\stackrel{(n)}{f_{\mu\nu}}\equiv\theta^{\alpha_1\beta_1}...\theta^{\alpha_n\beta_n}f_{\alpha_1\beta_1 ...\alpha_n\beta_n\mu\nu}
\end{equation}
(with $\stackrel{(0)}{f_{\mu\nu}}\equiv F_{\mu\nu}$). 

Using this notation the effective Lagrangian  (\ref{effNCYM}) is
\begin{equation*}
{L}^{eff}=-\frac{1}{4}\sum\limits_{k=0}^{\infty}\vartheta^{k}{L}^{eff(k)}, \qquad \mbox{where}\qquad {L}^{eff(k)}=\sum\limits_{i=0}^{k}\stackrel{(i)}{f_{\mu\nu}}\stackrel{(k-i)}{f^{\mu\nu}}.
\end{equation*}

All the terms in the effective Lagrangian are of the generic form, 
\begin{equation}
\label{f2}
\stackrel{(n)}{f_{\mu\nu}} \stackrel{(m)}{f^{\mu\nu}}, \qquad n,m\in \{0,1,2,... \quad | \quad n+m= k\},
\end{equation}
for $k=0,1,2,...$. Our program is to show that if this generic term is invariant under a global transformation (linear in $x^\mu$) the NCYM action (\ref{effNCYM}) will be invariant and the associated  transformation is a symmetry. In particular this will be the case for the Weyl transformations given in the previous section.

The Noether theorem can be written as, 
\begin{equation*}
\frac{\delta{L}}{\delta \phi^a}Q^a+\partial_\mu j^\mu=0.
\end{equation*}
where we want to implement the transformation $\phi^a \mapsto \phi^a + Q^a([\phi],x)$ as a symmetry of the Lagrangian ${L}([\phi],x)$ and  $j^\mu$ is the associated Noether current.
In the case of (\ref{effNCYM}) we have 
\begin{equation*}
\frac{\delta{L}^{ef}}{\delta A_\mu^A}Q^A_\mu+\partial_\mu j^\mu=0.
\end{equation*}
The transformation $Q^A_\mu$ can be written as a series in $\vartheta$ given by
\begin{equation*}
Q^A_\mu \equiv Q^{(0)A}_\mu+\vartheta Q^{(1)A}_\mu+\vartheta^2 Q^{(2)A}_\mu+\dots,
\end{equation*}
and the corresponding series for the conserved currents are
\begin{equation}
j^{\mu}=j^{(0)\mu}+\vartheta j^{(1)\mu}+\dots
\end{equation}
For each order $k$ in $\vartheta$ we have consistence conditions. To zero order in $\vartheta$ these equations imply the Weyl invariance of the usual Yang-Mills commutative action, with $Q^{(0)A}_\mu\equiv\delta_\xi A^A_\mu=-\LIE{\xi}{A^A_\mu}$ y $j^{(0)\mu}=-\xi^\mu{L}^{ef(0)}$ To first orden  in $\vartheta$ we have
\begin{equation}
\label{invriancia}
\frac{\delta{L}^{ef(1)}}{\delta A_\mu^A}\delta_\xi A^A_\mu+\frac{\delta{L}^{ef(0)}}{\delta A_\mu^A}Q^{(1)A}_\mu+\partial_\mu j^{(1)\mu}=0.
\end{equation}
with the trivial solution $Q^{(1)A}_\mu=0$  we can implement the transformation as a symmetry of the noncommutative action up to first order in $\vartheta$. 
A crucial observation is that all this woks thanks to the fact that we are using Weyl transformation and $\theta$ as a tensor.
In the same way we can calculate the associated Noether current $j^{(1)\mu}$ to first order in $\vartheta$ . 

A similar analysis was presented in \cite{BaBrGr} but taking  $\theta^{\mu\nu}$ as a matrix (not a tensor). 
The result of this authors is that the Poincar\'e symmetry is obstructed and only a subset of this symmetry can be implemented. The restriction over  $\theta$ is $\LIE{\xi}{\theta^{\mu\nu}}=0$. When this restriction is valid the transformation in question can be implemented  as a ``little global symmetry'' of the theory.  

As we will see our modified variational derivate  (\ref{twistvariacional}) can be implemented and is not obstructed. Applying it for the general term  (\ref{f}) and taking into account that $\partial_\sigma\xi^\rho$ is a constant and using $[\mathcal{L},\partial_\mu]=0$, we have
\begin{equation*}
W_\xi\stackrel{(n)}{f_{\mu\nu}} =-\sum\limits_{k=0}\partial_{\mu_1}...\partial_{\mu_k}\LIE{\xi}{A^A_\alpha}\frac{\partial \stackrel{(n)}{f_{\mu\nu}}}{\partial A^A_{\alpha,\mu_1\dots\mu_k}}-\frac12\LIE{\xi}{\theta^{\alpha\beta}}\frac{\partial \stackrel{(n)}{f_{\mu\nu}}}{\partial\theta^{\alpha\beta}}=-{\cal L}_\xi \stackrel{(n)}{f_{\mu\nu}}
\end{equation*}
($\partial A^A_{\alpha,\mu_1....\mu_k}\equiv\partial_{\mu_1}\dots\partial_{\mu_k} A^A_{\alpha}$). Using Killing relations (\ref{killingeq})  
we can write
\begin{equation*}
\delta_\xi(\stackrel{(n)}{f_{\mu\nu}}\stackrel{(m)}{f^{\mu\nu}})=-{\cal L}_\xi(\stackrel{(n)}{f_{\mu\nu}}\stackrel{(m)}{f^{\mu\nu}})=-\partial_\rho(\xi^\rho\stackrel{(n)}{f_{\mu\nu}}\stackrel{(m)}{f^{\mu\nu}}).
\end{equation*}
So we can conclude that the consistence conditions
 (\ref{invriancia}) to any order in $k$ and $\vartheta$ are satisfied for $Q_\mu^A=Q^{(0)A}_\mu=\delta_\xi A^A_\mu$ and then the Weyl symmetry of the noncommutative  Yang-Mills theory is NOT obstructed.

\section{Discussion: Covariance \textit{vs.} Invariance}

In this note we have presented an implementation of space-time symmetries in NCFT that in particular can be used to formulate Lorentz invariant noncommutative theories.
A central point of our analysis was to consider $\theta^{\mu\nu}$ as a two tensor in contrast with twist symmetries. In a second step we have showed that Weyl transformation leave the $\star$ product invariant. Then we have constructed a modified variational recipe (\ref{twistvariacional}) to implement  symmetries.  
This modification to the variational derivative was recognized by  \cite{BiGrGrKrPoScWu}, where the authors gives a simple by hand argument that works fine for the case of NCYM theory\footnote{The change in the noncommutative parameters is justified by making the observation that we can interpret in two different ways the implementation of a Lorentz transformation that correspond to active and passive transformations \cite{CaHaKoLaOk,BiGrGrKrPoScWu}.}.
Here we have constructed a universal modification applicable for any NCFT that was inspired in a physical motivated implementation of twist transformations.
 Our analysis can be justified also from the following perspective.

Given the transformation of the $\star$ product 
\begin{equation}
\phi'_1(x';\theta')\star'\phi'_2(x';\theta')=\left({\cal D}_1\phi_1(x;\theta)\right)\star\left({\cal D}_2\phi_2(x;\theta)\right), \label{transtens1}
\end{equation}
where $x^\mu\mapsto x'^\mu= x^\mu +\xi^\mu$ y $\theta^{\mu\nu}\mapsto\theta'^{\mu\nu}=\theta^{\mu\nu}+\delta\theta^{\mu\nu}$, we can define two types of infinitesimal transformations  \cite{Hi}
\begin{align*}
\gamma(\phi_1 \star \phi_2)&\equiv  \phi'_1(x';\theta')\star'\phi'_2(x';\theta') -\phi_1(x;\theta)\star \phi_2(x;\theta),   \\
\hat{\gamma}(\phi_1 \star \phi_2)&\equiv \phi'_1(x';\theta')\star'\phi'_2(x';\theta')-\phi_1(x';\theta')\star' \phi_2(x';\theta'),
\end{align*}
that in turn are related by
\begin{equation*}
\hat{\gamma}(\phi_1 \star \phi_2)=\gamma(\phi_1 \star \phi_2)-[\phi_1(x';\theta')\star' \phi_2(x';\theta')-\phi_1(x;\theta)\star \phi_2(x;\theta)].
\end{equation*}
In the particular case of  Weyl transformations  $\star=\star'$, so we have
\begin{multline}
\label{vargamgor}
\hat{\gamma}(\phi_1 \star \phi_2)=\gamma(\phi_1 \star \phi_2)-\Bigl[\phi_1(x;\theta)\star(\xi^\rho\partial_\rho \phi_2 +\frac12\delta\theta^{\alpha\beta}\frac{\partial \phi_2}{\partial\theta^{\alpha\beta}}) \\ +(\xi^\rho\partial_\rho \phi_1+\frac12\delta\theta^{\alpha\beta}\frac{\partial \phi_1}{\partial\theta^{\alpha\beta}})\star \phi_2(x;\theta)\Bigr],
\end{multline}

The  action of $\gamma$  for a product of two fields is now 
\begin{equation}
\label{acccov}
\hat{\gamma}(\phi_1 \star \phi_2)=\phi_1\star\left(\delta_{2,\xi} \phi_2-\frac{1}{2}\delta\theta^{\alpha\beta}\frac{\partial \phi_2}{\partial\theta^{\alpha\beta}}\right)+\left(\delta_{1,\xi} \phi_1-\frac{1}{2}\delta\theta^{\alpha\beta}\frac{\partial \phi_1}{\partial\theta^{\alpha\beta}}\right)\star \phi_2
\end{equation}
($\delta_{1(2),\xi}$ is the infinitesimal operator that correspond to $\mathcal{D}_{1(2)}$). As the variation $\delta\theta^{\mu\nu}\equiv-\LIE{\xi}{\theta^{\mu\nu}}$ and using (\ref{regmult}) we can identify the operator
\begin{equation}
\label{gamma}
\boxed{\hat\gamma=\delta_{\xi} -\frac12\delta\theta^{\alpha\beta}\frac{\partial }{\partial\theta^{\alpha\beta}}}.
\end{equation}
This operator {\it is} a derivation of the$\star$ product (fulfill  Leibniz rule). This is an important property of this differential operator but if $\delta_\xi$ is a symmetry of the theory is $\hat\gamma$ also a symmetry? If we take $\delta_\xi$ as $W_\xi$ from (\ref{twistvariacional}), then $\hat\gamma$ acts as a variational derivative (\ref{dervar}). But, as we know, the variational derivative is {\em not} a symmetry in NCFT.
If we take in (\ref{gamma})  $\delta_\xi$ as a twist symmetry, then $\delta_\xi\theta^{\mu\nu}=0$ and $\hat\gamma\theta^{\mu\nu}=-\delta\theta^{\mu\nu}$.  The operator (\ref{gamma}) with  $\delta_\xi$ as a twist symmetry was noticed by the authors of \cite{Gr-BoLiRuVi},  where they try to implement the Weyl symmetry in NCFT.  Following the construction of \cite{BiGrGrKrPoScWu} they implement the generators  $G_\xi=-\xi^\rho\partial_\rho$ from the standard Weyl group by adding a term proportional to the induced change $\theta$ and defining the new generators as
\begin{equation}
\label{grbogen}
G_\xi^\theta\equiv G_\xi -\frac12\delta_\xi\theta^{\alpha\beta}\frac{\partial }{\partial\theta^{\alpha\beta}}.
\end{equation} 
These generators satisfy the same algebra as the non deformed standard Weyl Lie algebra and act as derivations over the $\star$ product for functions of space-time that are also functions of $\theta$. In this way they extend the space of function of space-time to function of space-time and $\theta$, $f(x,\theta)$ on which the operators (\ref{grbogen}) can be implemented. This implementation is consistent for affine transformations. If we try to apply the same idea to general diffeomorphisms we find that the implementation of (\ref{grbogen}) as a symmetry is obstructed because they do not comply with the Leibniz rule.   

In contrast with the twist symmetries, the action of the generators  (\ref{grbogen}) is given by 
\begin{equation*}
m_\star(\Delta_\star(G)\triangleright f\otimes g) =G^\theta m_\star(f\otimes g)+\frac12\delta_\xi\theta^{\mu\nu}\frac{\partial}{\partial\theta^{\mu\nu}}m_\star(f\otimes g),
\end{equation*}
with $m_\star(f\otimes g)=f\star g$, and $\Delta_\star$ the deformed coproduct. 
From our perspective this result enforce the idea that the twist symmetries and the symmetries generated by
$\hat\gamma+\frac12\delta_\xi\theta^{\mu\nu}\frac{\partial}{\partial\theta^{\mu\nu}}$ (the analogous of  $G_\xi^\theta+\frac12\delta_\xi\theta^{\mu\nu}\frac{\partial}{\partial\theta^{\mu\nu}}=G_\xi$) with $\hat\gamma$ las the variational derivative (\ref{dervar}) give the same result when they act on the $\star$ product of two or more fields.  

In contrast with the result presented in \cite{Gr-BoLiRuVi}, we state that $G_\xi^\theta$, or $\hat\gamma$ presented in our analysis does not generate space-time symmetries since the action of $\hat\gamma$ as the variational derivative (\ref{dervar}) are derivation with respect to the $\star$ product. If we realize $\delta_\xi$ in  (\ref{gamma}) as a space-time symmetry  $\hat\gamma$ could be also a symmetry if $\frac{1}{2}\delta\theta^{\alpha\beta}\frac{\partial }{\partial\theta^{\alpha\beta}}$ is a divergence of some space-time function. But as we see, in general this is not the case. So we conclude that it is not possible to have at the same time space-time transformations that are symmetries of a given action in NCFT and derivation with respect to the $\star$ product.

\section*{Acknowledgements}

This work was partially sopported  by CONACyT grant 50-155I, as well as DGAPA-UNAM grant IN116408.

\end{document}